\documentclass[aip,jcp,reprint]{revtex4-1}

\usepackage{amsmath}
\usepackage{amsfonts}
\usepackage{url}
\usepackage{bm}
\usepackage{bbold}
\usepackage{graphics}
\usepackage{paralist}
\usepackage{graphicx}
\usepackage{afterpage}

\newcommand{\bra}{\langle}
\newcommand{\ket}{\rangle}
\newcommand{\bs}[1]{\boldsymbol{#1}}

\begin{document}

\title{Density matrices in full configuration interaction quantum Monte Carlo: Excited states, transition dipole moments and parallel distribution}

\author{N. S. Blunt}
\email{nicksblunt@gmail.com}
\affiliation{Max Planck Institute for Solid State Research, Heisenbergstra{\ss}e 1, 70569 Stuttgart, Germany}
\affiliation{University Chemical Laboratory, Lensfield Road, Cambridge, CB2 1EW, U.K.}
\author{George H. Booth}
\affiliation{Department of Physics, King's College London, Strand, London WC2R 2LS, U.K.}
\author{Ali Alavi}
\email{alavi@fkf.mpg.de}
\affiliation{Max Planck Institute for Solid State Research, Heisenbergstra{\ss}e 1, 70569 Stuttgart, Germany}
\affiliation{University Chemical Laboratory, Lensfield Road, Cambridge, CB2 1EW, U.K.}

\begin{abstract}
We present developments in the calculation of reduced density matrices (RDMs) in the full configuration interaction quantum Monte Carlo (FCIQMC) method. An efficient scheme is described to allow storage of RDMs across distributed memory, thereby allowing their calculation and storage in large basis sets. We demonstrate the calculation of RDMs for general states by using the recently-introduced excited-state FCIQMC approach [J. Chem. Phys.~\textbf{143}, 134117 (2015)] and further introduce calculation of transition density matrices (TDMs) in the method. These approaches are combined to calculate excited-state dipole and transition dipole moments for heteronuclear diatomic molecules, including LiH, BH and MgO, and initiator error is investigated in these quantities.
\end{abstract}

\maketitle

\section{Introduction}
\label{sec:intro}

The estimation of general expectation values in quantum Monte Carlo (QMC) methods is a challenging task, one that has been achieved with varying levels of success across the range of QMC methods employed and solutions attempted. In diffusion Monte Carlo (DMC)\cite{Foulkes2001}, where sampling is performed in real space, the energy can be estimated simply through the use of a mixed estimator, but approaches to calculate the expectation value of operators that do not commute with the Hamiltonian are considerably less straightforward. The estimation of such expectation values remains a difficult task in general\cite{Foulkes2001, Gaudoin2007, Casulleras1995, Russo2012}.

The full configuration interaction quantum Monte Carlo (FCIQMC) method\cite{Booth2009, Cleland2010, Booth2012} has much in common with DMC and other projector QMC Carlo methods. However, a crucial feature that separates FCIQMC from some such methods is that sampling is performed in a discrete basis set, usually comprised of basis states that share the antisymmetry of the exact fermionic wave function. While the use of a discrete basis set necessarily leads to a finite basis set error, it also leads to many of the methods favourable properties. Perhaps most notably, the discrete space allows for efficient annihilation that significantly ameliorates the sign problem\cite{Spencer2012, Kolodrubetz2013}. Furthermore, in a discrete space the so-called replica trick\cite{Zhang1993, Blunt2014} results in an incredibly simple and effective approach for estimating general observables.

The replica trick was first used in FCIQMC by Overy \emph{et al.}\cite{Overy2014} to stochastically sample the two-particle reduced density matrix in an unbiased manner. This was in contrast to previous efforts to the compute the 2-RDM, which suffered from large systematic errors\cite{Booth2012_2}. Replica-sampled 2-RDMs were subsequently used by Thomas \emph{et al.}, firstly to compute nuclear forces, dipole moments and polarizabilities for various small molecules\cite{Thomas2015_2}, and subsequently to perform orbital optimization in FCIQMC-based extensions to the multi-configurational self-consistent field approach, allowing the use of very large active spaces\cite{Thomas2015_3}. RDMs have also been fundamental to a recent study of explicitly correlated approaches in the FCIQMC method\cite{Kersten2016}. From the range of applications even at this early stage, it is clear that the calculation of RDMs in FCIQMC will be of key importance going forward.

There also exists a separate but important family of methods that construct the 2-RDMs without using the wave function\cite{rdm_book, Schlimgen2016, Mazziotti2016}. These methods instead rely on imposing N-representability conditions, with a lower bound on the energy improving as more conditions are enforced. In contrast, FCIQMC directly uses a stochastic sampling of the wave function to construct 2-RDMs, becoming exact as the initiator approximation\cite{Cleland2010, Cleland2011} is improved upon, and in the limit of longer sampling.

Current implementations of the FCIQMC algorithm make use of its stochastic nature to allow efficient large-scale parallelization. The walkers that sample the underlying wave function are distributed among processors, thus allowing the most substantial memory requirements to be distributed\cite{Booth2014}. Unfortunately, this has not been true of the reduced density matrices (RDMs), which in previously described implementations have been stored in their entirety on every processing core. This has limited the scope of applications thus far, especially with the emphasis on large active spaces that the method can effectively sample. 

Meanwhile, the current authors recently introduced an excited-state approach within the FCIQMC framework\cite{Blunt2015_3}. In this approach, multiple FCIQMC simulations are performed simultaneously, one for each state to be sampled. Simulations representing higher-energy states are orthogonalized against the stochastic snapshots of wave functions in simulations representing lower-energy states, thus preventing collapse to the ground state. It was shown that this approach appears to be possible without incurring detectable systematic errors. This therefore provides a very simple and accurate approach to sampling excited states, which we believe shows significant promise.

In this article we build upon these recent developments by introducing an extension to the computation of unbiased excited-state reduced density matrices and transition density matrices (TDMs). These latter objects are required in order to calculate transition amplitudes induced by different operators between excited states (such as an optical excitation), and critical in the connection of results to spectroscopic experiments. In order to circumvent the limiting storage of these sampled objects, we will furthermore describe an efficient approach to distribute such objects efficiently across distributed memory architecture.

In Section II we provide a brief overview of the excited-state FCIQMC method and of the calculation of RDMs within the ground-state FCIQMC approach. In Section III, the theory underlying the unbiased estimation of excited-state RDMs and TDMs is presented. Section IV then describes an efficient parallelized implementation of RDMs within a large-scale FCIQMC code, and results are presented in section V.

\section{Background theory}
\label{sec:theory}

\subsection{The excited-state FCIQMC method}

We begin with a brief overview of the recently-introduced excited-state FCIQMC method\cite{Blunt2015_3}. The ground-state FCIQMC algorithm is a projector QMC method that achieves a stochastic sampling of the ground-state wave function. This is accomplished by repeated application of a projection operator, $\hat{P}$, to some initial wave function. In FCIQMC, $\hat{P}$ is defined by 
\begin{equation}
\hat{P} = \mathbb{1} - \Delta \tau (\hat{H} - S \mathbb{1}),
\end{equation}
where $\hat{H}$ is the Hamiltonian, $\Delta \tau$ is a small imaginary-time step and $S$ is a diagonal shift parameter, varied slowly throughout the simulation to allow some control over the wave function normalization. We work in a discrete basis set, usually of Slater determinants, $\{ |D_i \ket \}$, and the projection operator can then be expressed as $\hat{P} = \sum_{ij} P_{ij} | i \ket \bra j |$. The FCIQMC method then primarily consists of an algorithm to sample elements of $\bs{P} \bs{\psi}$ (and through iteration therefore of $(\bs{P})^m \bs{\psi}$), for some initial $\bs{\psi}$, in an extremely efficient and unbiased manner. In the limit of large $m$, a stochastic sampling of the ground state is achieved. Detailed introductions to the FCIQMC algorithm, by which this is achieved, exist elsewhere\cite{Booth2009, Spencer2012, Booth2014}.

The excited-state FCIQMC method consists of a small extension to the ground-state approach, which prevents collapse to the ground-state through a simple orthogonalization procedure. Suppose the exact ground-state wave function was known, and denoted $|\Phi^0\ket$. Then the first excited-state could be converged upon by repeated application of the operator
\begin{equation}
\hat{P}_{\textrm{first excited}} = [\mathbb{1} - | \Phi^0 \ket \bra \Phi^0 |] \; \hat{P}.
\end{equation}
A single iteration of a QMC algorithm for the first-excited state could then be to apply $\hat{P}$ via FCIQMC's spawning rules, followed by applying $\mathbb{1} - | \Phi^0 \ket \bra \Phi^0 |$ to the resulting wave function. This latter projection exactly removes the ground-state contribution, and the next-lowest energy state would be converged upon instead. In practice this is not possible because the exact ground state is unknown. However, the basic FCIQMC method provides a procedure to stochastically sample this state. Therefore, by performing two FCIQMC simulations simultaneously, the first-excited state may be sampled in the second simulation by orthogonalizing against the instantaneous FCIQMC-sampled ground-state in the first simulation, a simple task to perform computationally. While one may worry that orthogonalizing against stochastic snapshots of a state is not equivalent to orthogonalizing against an exact wave function, we have been unable to detect any discrepancy whatsoever resulting from this difference, and this simple procedure appears unbiased beyond any achievable level of statistical accuracy\cite{Blunt2015_3}.

The excited-state FCIQMC algorithm to sample the lowest eigenstates of $\hat{H}$ can therefore be summarized as follows. Perform multiple simulations simultaneously, one for each state to be sampled. Let the (unnormalized) wave function sampled by simulation $n$ and at imaginary-time $\tau$ be denoted $|\Psi^{n}(\tau)\ket$. Then the evolution sampled by the various simulations is given by
\begin{equation}
|\Psi^{n}(\tau+\Delta\tau)\ket = \hat{O}^{n}(\tau+\Delta\tau) \; \hat{P} \; |\Psi^{n}(\tau)\ket,
\end{equation}
where
\begin{equation}
\hat{O}^{n}(\tau) = \mathbb{1} - \sum_{m < n} \frac{|\Psi^{m}(\tau)\ket \bra \Psi^{m}(\tau)|}{\bra \Psi^{m}(\tau) | \Psi^{m}(\tau) \ket}.
\label{eq:orthog_proj}
\end{equation}
A single iteration therefore firstly consists of applying the FCIQMC ground-state projection operator $\hat{P}$ to each simulation, which is achieved by the usual FCIQMC algorithm, and includes performing all annihilation. Only then is orthogonalization performed, with each simulation being orthogonalized against \emph{all lower-energy} states. This fully describes the excited-state adaptation to the FCIQMC algorithm, which can be performed at essentially linear cost with the number of desired states, since the orthogonalization step requires close to negligible computational overhead.

\subsection{Unbiased RDM estimation in FCIQMC}
\label{sec:rdm_estimation}

The second-order (symmetric) reduced density matrix (2-RDM) for a state $|\Phi^n\ket$ is defined as
\begin{equation}
\Gamma_{pq,rs}^n = \bra \Phi^n | a_p^{\dagger} a_q^{\dagger} a_s a_r | \Phi^n \ket,
\end{equation}
where $p$, $q$, $r$ and $s$ denote spin-orbital labels. In the rest of this article we will only be concerned with states representing the eigenstates of the Hamiltonian, $\hat{H}$. We therefore let $|\Phi^n\ket$ for $n>0$ denote the \emph{exact} $n$'th excited state of $\hat{H}$, while $|\Phi^0\ket$ denotes the exact ground state, and the determinant expansion of a state $|\Phi^n\ket$ is expressed as
\begin{equation}
| \Phi^n \ket = \sum_i \phi_i^n |D_i\ket.
\end{equation}
$\phi_i^n$ will be assumed to be real throughout, although complex FCIQMC extensions are straightforward\cite{Booth2012}. The RDM for energy eigenstate $|\Phi^n\ket$ can then be expressed as
\begin{equation}
\Gamma_{pq,rs}^n = \sum_{ij} \phi_i^n \phi_j^n \bra D_i | a_p^{\dagger} a_q^{\dagger} a_s a_r | D_j \ket.
\end{equation}

Let FCIQMC coefficients be denoted $C_i^n$. Then the FCIQMC wave function for state $n$ is $|\Psi^n \ket = \sum_i C_i^n | D_i \ket$, with $C_i^n$ denoting the (signed) number of walkers on determinant $|D_i\ket$. The FCIQMC and excited-state FCIQMC algorithms are unbiased at large enough walker numbers to a very good approximation, meaning that the average value of $C_i^n$ throughout the FCIQMC simulation should approach the exact coefficient, $\phi_i^n$. A more precise statement is that, once convergence has been reached,
\begin{equation}
E[C_i^n] = \phi_i^n,
\end{equation}
where $E[...]$ denotes the expectation value, equivalent to the long time-averaged walker population. We ignore for now that the FCIQMC wave function is not normalized.

A simple approach to sample an (unnormalized) RDM within FCIQMC would be through
\begin{equation}
\Gamma_{pq,rs}^n \stackrel{?}{\approx} \sum_{ij} C_i^n C_j^n \bra D_i | a_p^{\dagger} a_q^{\dagger} a_s a_r | D_j \ket.
\label{eq:biased_rdm}
\end{equation}
However, some thought quickly shows that this is not correct: the expectation of this expression does not give back the desired exact 2-RDM, because $E[C_i^n C_j^n] \ne \phi_i^n \phi_j^n$ unless $C_i^n$ and $C_j^n$ are uncorrelated with a zero covariance, which they will not be. This is particularly clear for diagonal elements: expressing $C_i^n = \phi_i^n + \delta_i^n$, then $E[(C_i^n)^2] = (\phi_i^n)^2 + (\delta_i^n)^2$, which is always greater than the desired exact value for a non-zero $\delta_i^n$. Therefore, averaging this expression over many iterations will result in a biased 2-RDM estimate. If one could average all FCIQMC coefficients over the entire simulation then the associated errors, $\delta_i^n$, could be reduced to a point where any bias is negligible, but this is not feasible due to the memory requirements of storing all averaged coefficients rather than the far smaller set of instantaneously occupied determinants. In an earlier implementation, Booth \emph{et al.} attempted a scheme where the FCIQMC coefficients are partially averaged over a reduced number of iterations, but an unacceptable bias was still found to occur\cite{Booth2012_2}.

This problem can be resolved by a replica sampling approach, whereby a second identical FCIQMC simulation is performed concurrently, starting from separate random number generator (RNG) seeds. The two FCIQMC simulations are performed for each state entirely independently, guaranteeing that the two wave function estimates will be uncorrelated. Let coefficients from the first and second replica simulations be denoted by $C_i^{1,n}$ and $C_i^{2,n}$, respectively. Then because the two simulations are uncorrelated, $E[C_i^{1,n} C_j^{2,n}] = \phi_i^n \phi_j^n$, and so
\begin{equation}
\Gamma_{pq,rs}^n = E \Big[ \; \sum_{ij} C_i^{1,n} C_j^{2,n} \bra D_i | a_p^{\dagger} a_q^{\dagger} a_s a_r | D_j \ket \; \Big].
\label{eq:fciqmc_rdm}
\end{equation}
While storing and averaging all coefficients in the FCIQMC wave function is not feasible, averaging all elements of $\Gamma_{pq,rs}^n$ is usually much more reasonable. In our previous implementation, we have therefore stored the entire 2-RDM in memory for each parallel process in a 2-dimensional non-sparse array. While possible for many small single-particle basis sets, this quickly becomes unreasonable, and Section~\ref{sec:implementation} is therefore dedicated to introducing a sparse and distributed implementation. We lastly remark that the above estimate for the 2-RDM in Eq.~\ref{eq:fciqmc_rdm} can be normalized in a straightforward fashion at the end of a calculation by enforcing that the trace of the 2-RDM must be equal to $N(N-1)/2$, as necessarily found in the exact 2-RDM.

\subsection{Practical considerations for sampling RDMs}
\label{sec:rdm_practical}

In this section, we describe some additional details and efficiency improvements regarding the practical calculation of 2-RDMs in FCIQMC which are relevant for the discussion on their distributed storage in Section~\ref{sec:implementation}. More details on many of these practical aspects can be found in Ref.~\onlinecite{Overy2014}. 

In general, we choose not to explicitly include all terms in Eq.~(\ref{eq:fciqmc_rdm}) exactly. Doing so would involve an expensive $\mathcal{O}(N^2 M^2)$ operation for each occupied determinant, where $N$ is the number of electrons and $M$ is the number of spatial orbitals. Instead we exploit the fact that the FCIQMC spawning procedure already involves randomly generating and sampling from the distribution of up-to-double excitations, and that $\{|D_i\ket, |D_j\ket\}$ pairs also only contribute to Eq.~(\ref{eq:fciqmc_rdm}) if within a double excitation of each other. The FCIQMC spawning procedure is therefore used to stochastically sample contributions to $\Gamma_{pq,rs}^n$, with contributions carefully weighted to avoid biases.

However, spawning attempts in FCIQMC are not performed for diagonal elements. Therefore, contributions to diagonal elements of the 2-RDM,
\begin{equation}
\Gamma_{pq,pq}^n = E \Big[ \sum_{\{p,q\} \in i} C_i^{1,n} C_i^{2,n} \Big],
\end{equation}
are added in exactly and explicitly. The summation here is performed over all determinants for which spin-orbitals $p$ and $q$ are both occupied. As a result of Brillouin's theorem, single excitations from the Hartree--Fock determinant in a canonical basis are also never generated, and all contributions to $\Gamma_{pq,rs}^n$ involving the Hartree--Fock determinant are therefore included explicitly. Doing this also greatly reduces stochastic noise in the 2-RDM estimate, since the Hartree--Fock determinant usually has a large amplitude. While it is possible that zero Hamiltonian matrix elements could result in the simulation `missing' further contributions to the RDM sampling, we have not found this to be a problem in practice.

Finally we note that instead of simply adding in contributions to the 2-RDM at every iteration, a more sophisticated block-averaging scheme is used. In this, the FCIQMC coefficients for the occupied determinants are averaged over ``blocks'' of many iterations. Contributions to diagonal elements of the 2-RDM and contributions involving the Hartree--Fock determinant are only added in at the end of each averaging block, using the FCIQMC coefficients averaged over the block length. All other contributions are added in on each iteration. This block averaging scheme was originally devised to help reduce biases in Eq.~(\ref{eq:biased_rdm}) before the replica sampling was in use\cite{Booth2012_2}, as discussed above. However, it also significantly reduces the computational cost associated with calculating diagonal and Hartree--Fock contributions to $\Gamma_{pq,rs}^n$, since these contributions are now only calculated infrequently. The calculation of diagonal elements is particularly expensive, with an $\mathcal{O}(N^2)$ cost per determinant, and so this averaging scheme is greatly beneficial computationally and provides identical results, despite not being required to avoid biased sampling when using the replica approach.

\section{Excited-state and transition density matrices}
\label{sec:excited_and_trdms}

\subsection{Excited-state density matrices}
\label{sec:excited_rdms}

The calculation of excited-state RDMs ($\Gamma_{pq,rs}^n$ with $n > 0$) in FCIQMC can be performed in exactly the same manner as for the ground state. The RDM estimator, Eq.~(\ref{eq:fciqmc_rdm}), depends only on the walker coefficients, $\bs{C}^{1,n}$ and $\bs{C}^{2,n}$. Coefficients for $n>0$ are stored and processed in exactly the same way as for $n=0$, and RDM sampling is therefore unchanged.

Two replica simulations must now be performed for each state being sampled, whereas the initial presentation of excited-state FCIQMC used only one such simulation for each state. However, this does not cause any difficulties beyond the additional computational effort required. One effectively performs two excited-state FCIQMC simulations concurrently, with walker coefficients in the first labelled $\bs{C}^{1,n}$ and those in the second labelled $\bs{C}^{2,n}$, as above. Orthogonalization is then performed equally in both simulations, i.e. via the operators
\begin{equation}
\bs{O}^{R,n} = \bs{I} - \sum_{m < n} \frac{ \bs{C}^{R,m} (\bs{C}^{R,m})^{\dagger} }{ (\bs{C}^{R,m})^{\dagger} \bs{C}^{R,m} }, \; \; \; R \in \{ 1, 2 \},
\label{eq:orthog_proj_rdm}
\end{equation}
with $R$ denoting the replica label, and $n$ denoting the state to which the operator is applied in the excited-state FCIQMC procedure. This choice ensures that all simulations with $R=1$ are uncorrelated from all those with $R=2$, and therefore avoids sampling biases in the RDM estimates, exactly analogously to the ground-state RDM estimation in Eq.~(\ref{eq:fciqmc_rdm}).
We note that one could modify the orthogonalization operator to make use of both replicas as follows,
\begin{equation}
\bs{O}^{n} \stackrel{?}{=} \bs{I} - \sum_{m < n} \frac{ \bs{C}^{1,m} (\bs{C}^{2,m})^{\dagger} }{ (\bs{C}^{1,m})^{\dagger} \bs{C}^{2,m} }.
\end{equation}
The motivation behind this is the desire to remove a theoretical non-linear bias in the operator, since $ E[ (\bs{C}^{1,m})^{\dagger} \bs{C}^{2,m} ] = \bra \Phi^{m} | \Phi^{m} \ket$, while $ E[ (\bs{C}^{1,m})^{\dagger} \bs{C}^{1,m} ] \ne \bra \Phi^{m} | \Phi^{m} \ket$. However, in practice this approach introduces correlations between walkers from the two replicas that biases RDM estimates, and as noted previously, a bias from the operator in Eq.~(\ref{eq:orthog_proj_rdm}) has not been detected in practice.

\subsection{Transition density matrices}
\label{sec:trdms}

A two-body transition density matrix (TDM) is defined by
\begin{equation}
\Gamma_{pq,rs}^{nm} = \bra \Phi^m | a_p^{\dagger} a_q^{\dagger} a_s a_r | \Phi^n \ket,
\end{equation}
and the equivalent one-body TDM is defined by
\begin{align}
\gamma_{p,q}^{nm} &= \bra \Phi^m | a_p^{\dagger} a_q | \Phi^n \ket, \\
                               &= \frac{1}{N-1} \sum_a \Gamma_{pa,qa}^{nm}.
\end{align}
These objects are of significance in spectroscopy and other processes where transitions are induced between eigenstates due to some operator. For instance, optical transitions are driven by the electric field, and therefore the transition dipole moment between states $m$ and $n$ can be calculated as
\begin{equation}
\bs{t}_{nm} = \sum_{pq} \gamma_{p,q}^{nm} \bra p | \hat{\bs{r}} | q \ket,
\label{eq:trans_dip_mom}
\end{equation}
where $| p \ket$ denotes a single-particle spin orbital and $\hat{\bs{r}}$ denotes the position operator. From this, using Fermi's golden rule, transition probabilities can be obtained. Note that the TDM (and hence e.g. transition dipole moment) is only determined up to an arbitrary sign, and physical observables must depend on the square modulus of this quantity.

One may suppose that TDMs can be estimated using similar estimators as introduced above, by
\begin{equation}
\Gamma_{pq,rs}^{nm} = E \Big[ \; \sum_{ij} C_i^{1,m} C_j^{2,n} \bra D_i | a_p^{\dagger} a_q^{\dagger} a_s a_r | D_j \ket \; \Big].
\end{equation}
or equivalently by
\begin{equation}
\Gamma_{pq,rs}^{nm} = E \Big[ \; \sum_{ij} C_i^{2,m} C_j^{1,n} \bra D_i | a_p^{\dagger} a_q^{\dagger} a_s a_r | D_j \ket \; \Big].
\end{equation}

However, because the FCIQMC wave functions are not normalized, the above objects also require explicit normalization. Non-transition (symmetric) 2-RDMs are normalized by noting that their trace must equal $N(N-1)/2$. TDMs are not so simple to correct, since their traces equal $0$ due to the orthogonality of the states. Instead they must be normalized directly by calculating the norm of the FCIQMC wave functions, which should equal $1$.

Defining the normalization of FCIQMC wave function by $E[| \Psi^{R,m} \ket ] = A^{R,m} |\Phi^{R,m} \ket$, it is anticipated that one way that the normalization of the TDMs could occur is with the following scheme,
\begin{align}
\label{eq:trdm_1}
\Gamma_{pq,rs}^{nm}[1] &\equiv \frac{E \Big[ \; \bra \Psi^{1,m} | a_p^{\dagger} a_q^{\dagger} a_s a_r | \Psi^{2,n} \ket \; \Big]}{ \sqrt{ E[ \bra \Psi^{1,m} | \Psi^{2,m} \ket ] E[ \bra \Psi^{1,n} | \Psi^{2,n} \ket ] }} \\
  &= \frac{ A^{1,m} A^{2,n} }{ \sqrt{ A^{1,m} A^{2,m} A^{1,n} A^{2,n} } } \; \Gamma_{pq,rs}^{nm}.
\end{align}
\begin{align}
\label{eq:trdm_2}
\Gamma_{pq,rs}^{nm}[2] &\equiv \frac{E \Big[ \; \bra \Psi^{2,m} | a_p^{\dagger} a_q^{\dagger} a_s a_r | \Psi^{1,n} \ket \; \Big]}{ \sqrt{ E[ \bra \Psi^{1,m} | \Psi^{2,m} \ket ] E[ \bra \Psi^{1,n} | \Psi^{2,n} \ket ] }} \\
  &= \frac{ A^{2,m} A^{1,n} }{ \sqrt{ A^{1,m} A^{2,m} A^{1,n} A^{2,n} } } \; \Gamma_{pq,rs}^{nm}.
\end{align}
However, this would only be sufficient if both replicas are always normalized equally, $A^{1,m} = A^{2,m}, \forall m$, otherwise non-linear expressions of stochastic quantities would result in a bias in the results. In general, the normalization of FCIQMC simulations can only be controlled to a certain extent, and attempting to enforce a constant normalization too aggressively can lead to additional biases\cite{Vigor2015}. Relying on the assumption that $A^{1,m} \approx A^{2,m}$ would likely be unreliable, and we proceed with a different approach which avoids this.

The product of the above two expressions gives the square of the desired result:
\begin{equation}
\Gamma_{pq,rs}^{nm}[1] \times \Gamma_{pq,rs}^{nm}[2] = (\Gamma_{pq,rs}^{nm})^2.
\end{equation}
However, this expression is not practical for calculating individual elements of $\Gamma_{pq,rs}^{nm}$, since the required square root operation is highly non-linear and introduces biases. This is most debilitating for very small elements of $\Gamma_{pq,rs}^{nm}$, where the stochastic samples, $\Gamma_{pq,rs}^{nm}[1]$ and $\Gamma_{pq,rs}^{nm}[2]$ have large stochastic relative errors, and therefore can have opposite signs, resulting in imaginary components of the TDM.

We therefore do not attempt to estimate individual elements of $\Gamma_{pq,rs}^{nm}$, but instead consider the final desired physical estimates. Such estimates are calculated as traces of the underlying density matrices, such as $\textrm{Tr}(\hat{O} \; \hat{\Gamma}^{nm})$ for some operator $\hat{O}$. Because the trace is a linear operation,
\begin{equation}
\textrm{Tr}(\hat{O} \; \hat{\Gamma}^{nm}) = \sqrt{ \textrm{Tr}(\hat{O} \; \hat{\Gamma}^{nm}[1]) \times \textrm{Tr}(\hat{O} \; \hat{\Gamma}^{nm}[2]) }.
\label{eq:trdm_estimator}
\end{equation}

This estimator does not suffer from a serious non-linear bias, because only a single square root operation is taken once all averaging and all summations have been performed. This final quantity will have a small relative error. This is in contrast to the above approach where a square root operation is performed for \emph{every} element of the density matrix, many of which will have large relative errors. Furthermore, since observables correspond to squared contractions of the TDM, generally this final square root operation will not generally be required, and truly unbiased computation can be made directly for the square of the transition operator.

To perform this unbiased sampling according to Eq.~(\ref{eq:trdm_estimator}), two arrays must be stored and averaged for each TDM being calculated, with averaging being performed according to the expectation values in Eqs.~(\ref{eq:trdm_1}) and (\ref{eq:trdm_2}). Physical quantities, such as transition dipole moments, can then be calculated according to Eq.~(\ref{eq:trdm_estimator}), and errors estimated using the standard propagation of error formula. The algorithmic sampling of $\Gamma^{nm}_{pq,rs}[1]$ and $\Gamma^{nm}_{pq,rs}[2]$ is identical to that of $\Gamma^n_{pq,rs}$ for non-transition RDMs. This includes the use of an array to average walker coefficients over blocks, in order to avoid calculating expensive diagonal density matrix elements every iteration, as discussed in Section~\ref{sec:rdm_practical}. Because the boundaries of these blocks depend on \emph{both} FCIQMC simulations contributing to a density matrix (see Ref.~(\onlinecite{Overy2014}) for details), separately averaged walker coefficients must be stored for \emph{each} density matrix sampled. This increases memory requirements, but these requirements are usually not too severe in practice. If such memory requirements do become limiting, then it may be possible to devise a different block-averaging scheme that avoids this greater memory requirement.

\section{Parallel and sparse implementation of density matrix arrays}
\label{sec:implementation}

In previous implementations, 2-RDMs in FCIQMC simulations were stored in their entirety for each compute process, in non-sparse two-dimensional arrays. Even accounting for the symmetries $\Gamma_{ij,kl} = -\Gamma_{ji,kl} = -\Gamma_{ij,lk}$ and $\Gamma_{ii,kl} = \Gamma_{ij,kk} = 0$, and using spin symmetry, this still leaves a total of $(M(M-1)/2)^2$ elements to be held in total for each combination of spin labels ($\Gamma_{\alpha\alpha,\alpha\alpha}$, $\Gamma_{\alpha\beta,\alpha\beta}$ and $\Gamma_{\alpha\beta,\beta\alpha}$ for restricted Hartree--Fock orbital bases), where $M$ is the number of spatial orbitals. While feasible for small orbital basis sets, this quickly becomes unmanageable. For example, in Section~\ref{sec:results} we consider LiH and BH molecules in aug-cc-pVQZ basis sets, both of which contain $126$ spatial orbitals. Assuming 64-bit integers, this gives a total memory requirement of around $500$MB per process and per RDM desired. Noting that a separate array must be held for each excited state RDM calculated, and two for each TDM, as well as arrays for both instantaneous and averaged estimates when desired, this quickly becomes impossible.

One option is to output elements to a file on a disk drive, but this would be unacceptable due to high access times. Another option is to not attempt to store such density matrices, but rather add their contributions to estimators on-the-fly. However, there are many situations where density matrices are required for post-FCIQMC calculations\cite{Thomas2015_3, Kersten2016, Sharma2014}. We therefore now describe a parallel and sparse implementation, which we have implemented in our FCIQMC program, \url{NECI}\cite{NECI_github, Booth2014}. The approach taken is similar to the sparse and parallel implementation of the FCIQMC walker array, already in use and described previously\cite{Booth2014}.

\begin{figure}[t!]
  \includegraphics{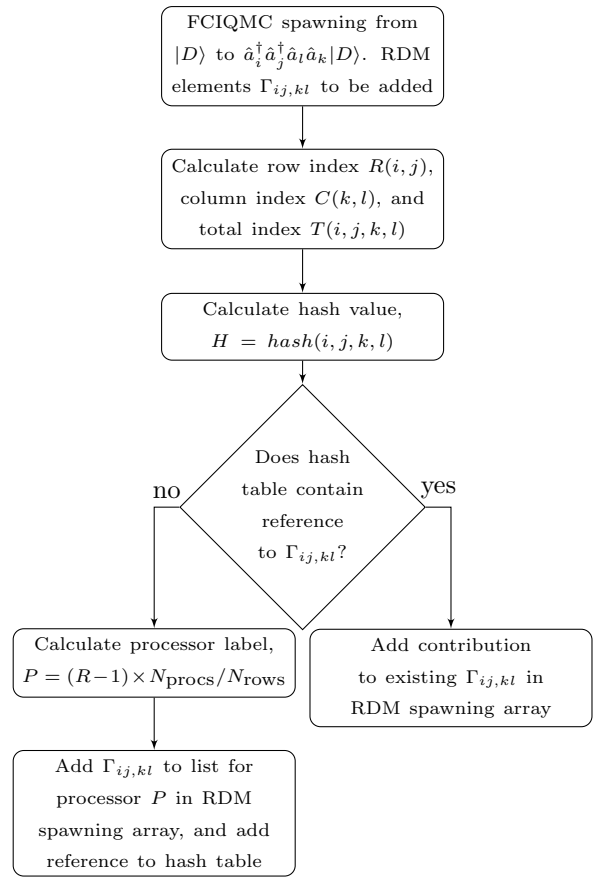}
  \caption{Algorithm used to add an RDM element to the RDM spawning array, after a double excitation occurs in the FCIQMC spawning step.}
  \label{fig:algorithm_1}
\end{figure}

We now choose to store all RDM elements in a sparse two-dimensional array. This array contains all density matrices simultaneously, both for the ground state and excited state RDMs, and for all TDMs. The process by which sampled elements are added to this array over the course of a single iteration is described in Figure~\ref{fig:algorithm_1}. Firstly, a double excitation is generated by the FCIQMC spawning algorithm, from a determinant $|D\ket$ to $\hat{a}^{\dagger}_i \hat{a}^{\dagger}_j \hat{a}_l \hat{a}_k |D\ket$, which leads to a sampling of a 2-RDM element $\Gamma_{ij,kl}$, as described in Section~\ref{sec:rdm_estimation}. In practice, a contribution is calculated for every density matrix involving the state from which FCIQMC spawning occurred. Thus, a single FCIQMC spawning leads a contribution in multiple density matrix samples.

Because a sparse implementation is desired, it is necessary to store the spin orbital labels, $i$, $j$, $k$ and $l$, in addition to the value $\Gamma_{ij,kl}$. To this end, we define a row label $R(i,j)$, a column label $C(k,l)$, and a total combined label $T(i,j,k,l)$ according to
\begin{align}
R(i,j) &= (i-1) M + j, \\
C(k,l) &= (k-1) M + l, \\
T(i,j,k,l) &= (R(i,j)-1) M^2 + C(k,l).
\end{align}
To save memory, the total label, $T(i,j,k,l)$, is stored as a single integer in the array of RDM contributions, rather than the individual spin orbital labels. This encoding does not make use of all symmetries but is efficient to decode, allowing individual orbital labels to be accessed when required. For a single-particle basis with $M$ spin orbitals, the largest 2-RDM contribution index is, $T(M,M,M,M) = M^4$. This allows indexing for an $M$ of more than $10^4$, far larger than any FCIQMC calculation to date.

Importantly, we note that certain RDM elements are generated very many times in an iteration. It is clearly inefficient and memory intensive to store these repeated elements separately throughout the spawned RDM array. We therefore ensure that repeated elements are always added into the same position. To achieve this, an efficient method is required to check if an element already exists in the array, and if so, where. We therefore also store a hash table to the RDM array. When a 2-RDM element $\Gamma_{ij,kl}$ is generated, a hash function is applied to $i$, $j$, $k$ and $l$, to calculate an associated hash value (the hash function applied is the same one used for the parallel distribution of determinants in FCIQMC, as presented in Ref.~(\onlinecite{Booth2014})). This hash value specifies which row of the hash table in which to search for this RDM element. If the element already exists in the RDM array, then a reference to that element will exist in that row of the hash table, and so the position of the existing element may be quickly looked up. This avoids having to perform expensive and regular sort operations on the RDM array to prevent the proliferation of repeated elements. In practice, each ``row'' of the hash table is a linked list implemented with pointers, whose size can increase dynamically as more hash clashes occur.

A mapping is also required to distribute density matrix elements across process threads within a distributed memory. In FCIQMC implementations, determinants have been distributed using a hash function of their occupied orbital list, to attempt to evenly distribute walkers across processors, balancing the load. Here, we simply split all possible RDM elements across processes evenly using the row label, $R(i,j)$. We find this adequate for calculations performed thus far, and it makes the printing of RDM arrays in a predefined order (as required in by many quantum chemistry packages) efficient and simple.

Finally, we consider efficient parallel distribution of the memory requirements. For a simulation performed on $N_{\textrm{proc}}$ processors, the total RDM spawning array is split evenly into $N_{\textrm{proc}}$ sections. When an RDM element is generated, its processor label is calculated, and the element is simply added into the earliest available position for the assigned process. In this way, it is then simple to distribute RDM elements to their correct processors at the end of an iteration via an \url{MPI_Alltoallv}, when using a Message Passing Interface (MPI) implementation. Before communication, RDM elements belonging to any MPI process can be held by any other process. However, for large RDMs, only a small fraction of elements are sampled in any one iteration, and so the memory requirements remain small. Then, once elements have been sent to their assigned processes, they can be averaged and accumulated over the entire simulation, with the greater memory requirements of this averaging efficiently distributed across processing cores.

Beyond allowing RDM arrays to be distributed across processing cores, there are further benefits to this approach. In particular, the full sparsity of the exact 2-RDMs is automatically taken advantage of entirely for a completely general system. This is because RDM elements are only added to the array once sampled, and only non-zero elements are ever sampled. Indeed, this will also be true for higher-order density matrices, including the full $N$-body density matrix, if desired. In contrast, taking full advantage of symmetries in the previously used scheme required separate computer code for separate systems, which often became difficult to maintain.

There are undoubtedly some overheads in this scheme. In particular, the fact that a hash lookup is performed for each RDM element sampled can be expensive. This is especially true since our implementation of the hash table is built from linked lists, which are slow to search compared to contiguous memory lookups. However, as described in Section~(\ref{sec:rdm_practical}), a block averaging scheme is used where diagonal elements, and elements involving the reference, are calculated very infrequently. These account for the majority of RDM elements sampled in practice, and so long as these elements are included infrequently, the overhead is small. There are also some memory overheads, in particular the requirement to store a hash table for each RDM array. However, this is easily outweighed by the parallel distribution and automatic exploitation of sparsity in all matrices. This is demonstrated in the next section, where 2-RDMs are sampled in single-particle basis sets which would have not been possible in the previous scheme. 

\section{Results}
\label{sec:results}

As an initial test of these ideas, we consider the calculation of dipole moments, transition dipole moments, and oscillator strengths for low-lying states of small diatomic molecules. These quantities are of great importance for understanding various properties of molecular systems. The oscillator strength in particular is required to explain optical spectra, as it determines the probabilities of absorption and emission of photons coupling different electronic states. Nonetheless, dipole moments are challenging to calculate accurately, even for small molecules, because they are very sensitive to the quality of the wave function and single-particle basis set used, generally requiring many diffuse orbitals for an accurate description, with far greater basis set sensitivity than the energy\cite{Green1974}.

We therefore begin by considering the LiH and BH molecules in aug-cc-pVDZ, aug-cc-pVTZ and aug-cc-pVQZ, containing $32$, $69$  and $126$ spatial orbitals respectively. The aug-cc-pVQZ basis 2-RDM was unobtainable in the previous RDM implementation, despite the small molecular size. We then consider the MgO molecule in an aug-cc-pVDZ basis set. We note that while the calculation of dipole moments only requires the 1-RDM, for these calculations we obtain the 1-RDM by contracting the 2-RDM, which we also use to calculate the energy using the estimator
\begin{equation}
(E_{\textrm{RDM}})_n = \frac{ \textrm{Tr} \big[ \hat{H} \; \hat{\Gamma}^n \big] }{ \textrm{Tr} \big[ \hat{\Gamma}^n \big] }.
\label{eq:rdm_energy}
\end{equation}
Therefore, the following is a good test of the newly-introduced ideas, as well as providing further insight into the effect of the initiator adaptation for different estimators and excited states.

The dipole moment for the state $|\Phi^n\ket$ is defined by
\begin{equation}
\bs{\mu}_{n} = \sum_{pq} \gamma_{p,q}^{n} \bra p | \hat{\bs{r}} | q \ket.
\end{equation}
while a transition dipole moment, $\bs{t}_{nm}$, is defined by Eq.~(\ref{eq:trans_dip_mom}), and the corresponding oscillator strength by
\begin{equation}
f_{nm} = \frac{2}{3} \Delta E_{nm} |\bs{t}_{nm}|^2,
\end{equation}
for an energy gap of $\Delta E_{nm}$ between states $|\Phi^n\ket$ and $|\Phi^m\ket$.

\begin{figure*}[t!]
\includegraphics{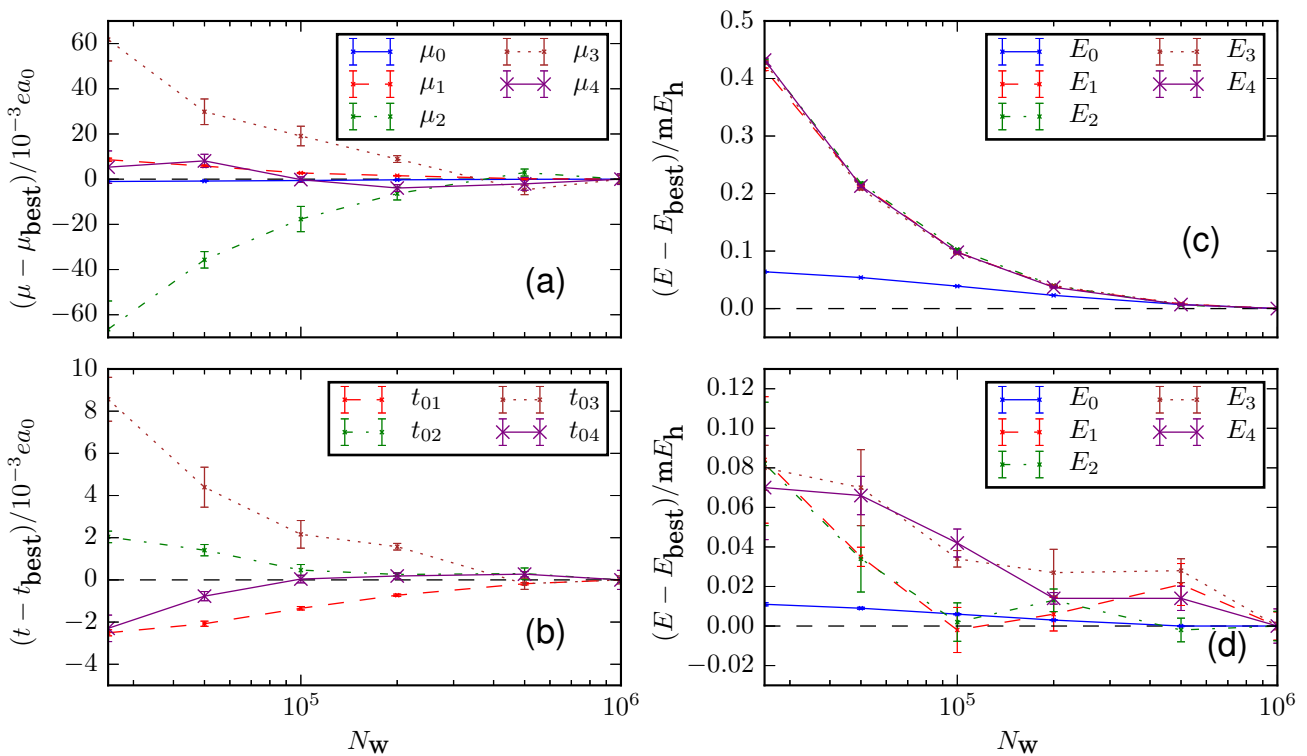}
\caption{Initiator error convergence for the five lowest energy states of LiH in an aug-cc-pVQZ basis, at an internuclear distance of $1.5957$\AA~as the number of walkers in each distribution is increased. Results are shifted relative to their values at the largest walker population considered, therefore approximately representing the initiator error. (a) Dipole moments. (b) Transition dipole moments from the ground state. (c) Energy calculated from a trial estimator, $E_{\textrm{Trial}}$. (d) Energy calculated from the RDM estimator, $E_{\textrm{RDM}}$. $N_w$ denotes the number of walkers for \emph{each} state and replica sampled. Simulations were typically averaged over 5 simulations to obtain error bars.}
\label{fig:lih_init}
\end{figure*}

\begin{table*}[t]
\begin{center}
{\footnotesize
\begin{tabular}{@{\extracolsep{4pt}}lccccc@{}}
\hline
\hline
Basis & State $n$ & Energy gap ($\Delta E_{0n}$) & Dipole moment ($\mu_n$) & Transition dipole moment ($t_{0n}$) & Oscillator strength ($f_{0n}$) \\
\hline
aug-cc-pVDZ & 0 $\;$ (${}^1\Sigma^+$)  & -            & -2.3251372(2) & -            & -           \\
            & 1 $\;$ (${}^1\Sigma^+$)  & 0.130434(1)  &  2.01947(4)   & 0.965189(7)  & 0.081007(1) \\
            & 2 $\;$ (${}^1\Sigma^+$)  & 0.2149799(6) & -3.3543(9)    & 0.37471(1)   & 0.020123(1) \\
            & 3 $\;$ (${}^1\Sigma^+$)  & 0.229077(4)  &  5.0832(8)    & 0.09126(8)   & 0.001271(2) \\
            & 4 $\;$ (${}^1\Sigma^+$)  & 0.246350(3)  & -0.2958(3)    & 0.56074(2)   & 0.051639(4) \\
\hline
aug-cc-pVTZ & 0 $\;$ (${}^1\Sigma^+$) & -            & -2.306440(9)  & -            & -           \\
            & 1 $\;$ (${}^1\Sigma^+$) & 0.132458(3)  &  2.02541(7)   & 0.93538(2)   & 0.077262(4) \\
            & 2 $\;$ (${}^1\Sigma^+$) & 0.216705(6)  & -3.794(1)     & 0.41146(2)   & 0.024459(2) \\
            & 3 $\;$ (${}^1\Sigma^+$) & 0.230621(2)  &  5.533(1)     & 0.07042(8)   & 0.000762(2) \\
            & 4 $\;$ (${}^1\Sigma^+$) & 0.246520(2)  & -0.6235(7)    & 0.693170(7)  & 0.078966(2) \\
\hline
aug-cc-pVQZ & 0 $\;$ (${}^1\Sigma^+$) & -            & -2.30168(3)   & -            & -           \\
            & 1 $\;$ (${}^1\Sigma^+$) & 0.132943(7)  &  2.0188(1)    & 0.92658(4)   & 0.076093(7) \\
            & 2 $\;$ (${}^1\Sigma^+$) & 0.217616(7)  & -3.696(2)     & 0.3984(1)    & 0.02303(1)  \\
            & 3 $\;$ (${}^1\Sigma^+$) & 0.231229(2)  &  6.211(2)     & 0.1083(2)    & 0.001809(6) \\
            & 4 $\;$ (${}^1\Sigma^+$) & 0.242846(9)  & -1.998(2)     & 0.6201(5)    & 0.06224(9)  \\
\hline
\hline
\end{tabular}
}
\caption{Final converged estimates for the LiH molecule at an internuclear distance of $1.5957$\AA. Results are for the five lowest energy states in the $A_1$ irrep of the $C_{2v}$ point group, with $M_S=0$ and $S=\textrm{even}$ quantum numbers (which happen to all be ${}^1\Sigma^+$ states). $n=0$ refers to the ground state, $n>1$ to excited states. Numbers in parentheses denote stochastic error, not initiator error. Energy gaps ($\Delta E_{1n}$) were calculated using RDM-based energy estimates, Eq.~(\ref{eq:rdm_energy}). Integrals were generated using the PySCF program\cite{pyscf}. In the small aug-cc-pVDZ, all results were verified against exact FCI results obtained from PySCF (not shown here).}
\label{tab:lih}
\end{center}
\end{table*}

For all simulations, the intial restricted Hartree--Fock (RHF) calculation was performed by PySCF\cite{pyscf}. Integrals from PySCF were then passed to our FCIQMC program, \url{NECI}, for the main calculation, which output one and two body density matrices. These were then contracted with integrals from PySCF to calculate final dipole moment estimates. Energy estimates were calculated on-the-fly in \url{NECI}.

The five lowest energy states were calculated for LiH and BH, and the four lowest states of MgO, considering only states with $M_s=0$ and using the $A_1$ irreducible representation (irrep) of the $C_{2v}$ point group. Also, time-reversal symmetrized functions\cite{Smeyers1973} were used as the many-particle basis states, therefore restricting the total spin quantum number, $S$, to be even, and thus removing triplet states. In all cases, the FCIQMC simulation time step was varied in the initial iterations so as to prevent ``bloom'' events, where many walkers can be created in a single spawning event (which often leads to large initiator error).

We also note that in generating excitations for the walker spawning step, we use an approach that greatly improves efficiency compared to the uniform sampling used in early FCIQMC results\cite{Booth2009}. In this approach, the pair of orbital labels from which electrons are excited, $(i,j)$, are chosen uniformly, while the orbitals excited to, $(a,b)$, are selected with probabilities drawn from a Cauchy-Schwarz distribution, namely $p(ab|ij) \propto \sqrt{\langle ia|ia \rangle \langle jb|jb \rangle}$.\cite{Smart_unpublished} Another approach to select connections efficiently was considered by Holmes \emph{et. al.}\cite{Holmes2016}, but not used here.

All simulations used the semi-stochastic adaptation to reduce stochastic errors\cite{Petruzielo2012, Blunt2015}. For the LiH molecule the deterministic space consisted of all configurations up to and including double excitations from the Hartree--Fock determinant. For the BH and MgO molecules the deterministic space was formed from the $10^4$ most populated configurations across all wave functions sampled, once the simulations were deemed to have largely converged, using the approach described in Ref.~(\onlinecite{Blunt2015}).

\subsection{LiH}

Simulations on LiH were performed using between $1.25 \times 10^4$ and $10^6$ walkers per simulation (i.e., for each state and replica sampled), in order to converge initiator error for all states. Density matrices were typically averaged over $10^5$ iterations, once convergence was deemed to have been reached for all states and all estimators. These entire simulations were then repeated five times with different initial RNG seeds, and the results averaged in order to calculate error estimates.

Figure~\ref{fig:lih_init} shows initiator convergence for LiH in the aug-cc-pVQZ basis set, for the lowest five energy eigenstates, and for four different estimators: dipole moments, transition dipoles moments, and energies calculated from both the RDM-based energy estimator, Eq.~(\ref{eq:rdm_energy}), and from a trial wave function-projected estimator:
\begin{equation}
(E_{\textrm{Trial}})_n = \frac{ \bra \Psi_{\textrm{Trial}}^n | \hat{H} | \Psi^n \ket }{ \bra \Psi_{\textrm{Trial}}^n | \Psi^n \ket }.
\label{eq:trial_energy}
\end{equation}
Here, $| \Psi_{\textrm{Trial}}^n \ket$ is a trial wave function designed to have a large overlap with the exact state $| \Phi^n \ket$. We have discussed the use of such trial wave function estimators in excited-state FCIQMC in Ref.~(\onlinecite{Blunt2015_3}). To generate $| \Psi_{\textrm{Trial}}^n \ket$, we calculate the configuration interaction singles and doubles (CISD) wave functions for the lowest fifteen energy states. Then, once convergence of all FCIQMC simulations is deemed to have been reached, we assign each simulation one trial wave function by choosing the CISD solution with the largest overlap in each case. The reason for obtaining more CISD solutions than FCIQMC simulations is that CISD solutions can have a different energy ordering to FCI solutions. Averaging of each $E_{\textrm{Trial}}$ estimate was performed from roughly the same point that RDM sampling began, and so both RDM and trial energy estimates are obtained from a similar number of iterations, usually $10^5$.

The initiator-FCIQMC estimates in Figure~\ref{fig:lih_init} are all plotted relative to their values at the largest walker population considered, $N_{w}=10^6$. Here, convergence has been largely reached in all cases, and so the figures effectively plot initiator error against walker population. Reassuringly, initiator error in energy estimates is incredibly small for both estimators and for all states. Indeed, the largest error at the smallest walker population tested is less than $\sim 0.5$ m$E_\textrm{h}$ for $E_{\textrm{Trial}}$.

Interestingly, initiator error in $E_{\textrm{RDM}}$ is much smaller than in $E_{\textrm{Trial}}$. This is a trend that we have often observed, although exceptions do occur (and in the limit of an exact $| \Psi_{\textrm{Trial}}^n \ket$, the initiator error is zero). Initiator error in the $E_{\textrm{RDM}}$ energies are variational in all cases within stochastic errors, while it is not strictly enforced (though common) for this to also be the case for $E_{\textrm{Trial}}$. For RDM-based energy estimates, this variationality is effectively ensured by the Hylleraas-Undheim-McDonald theorem\cite{Hylleraas1930, McDonald1933}, which is expected to approximately hold for FCIQMC-sampled wave functions. Initiator error is larger for excited states, as previously observed\cite{Blunt2015_3}. This is expected due to the more multi-configurational nature of excited states. It remains to be seen whether orbital optimization can increase this rate of convergence for excited states. Random errors however are larger in the RDM-based energy estimates, which is expected due to the fact that two uncorrelated simulations (from the two replicas) contribute to this quantity. However, error bars are extremely small in all cases here, always being smaller than $10^{-2}$ m$E_{\textrm{h}}$.

\begin{figure*}[t!]
\includegraphics{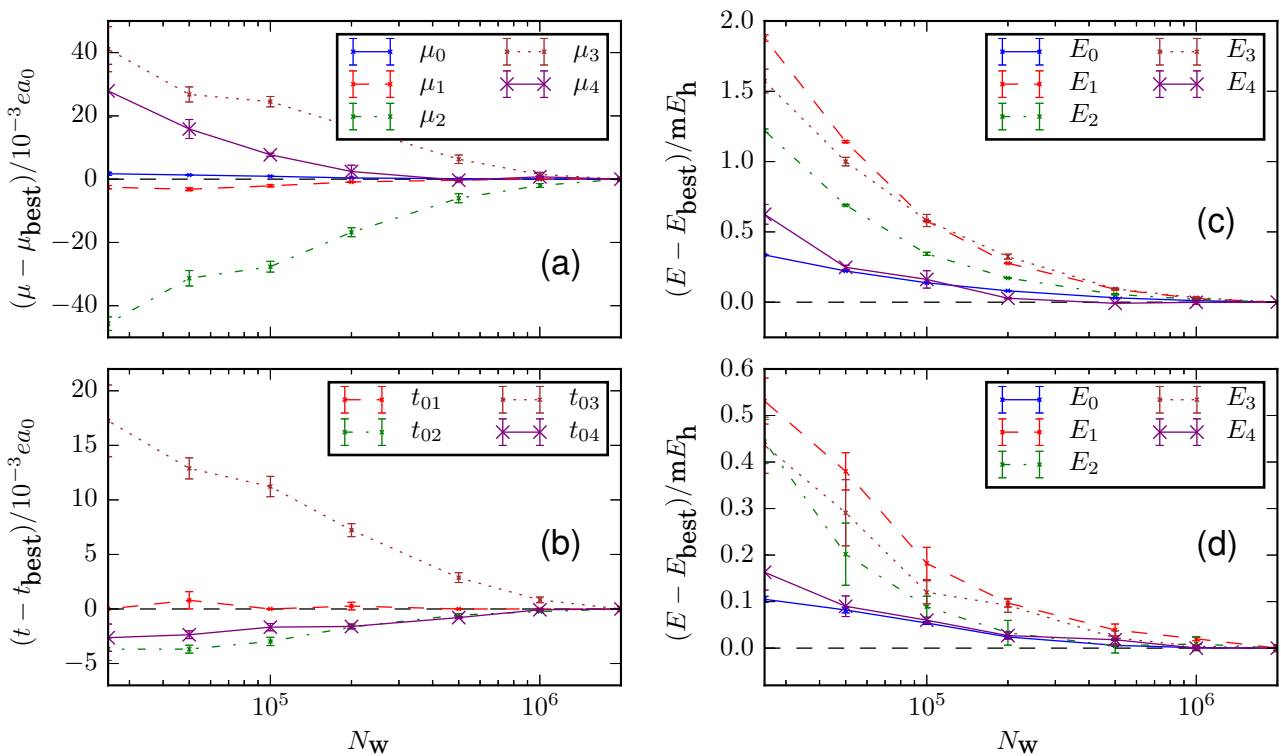}
\caption{Initiator convergence for the five lowest energy states of BH in an aug-cc-pVTZ basis, at an internuclear distance of $1.2324$\AA. Results are shifted relative to their values at the largest walker population considered, therefore approximately representing the initiator error. (a) Dipole moments. (b) Transition dipole moments from the ground state. (c) Energy calculated from a trial estimator, $E_{\textrm{Trial}}$. (d) Energy calculated from the RDM estimator, $E_{\textrm{RDM}}$. $N_w$ denotes the number of walkers for \emph{each} state and replica sampled. Simulations were typically averaged over 5 simulations to obtain error bars.}
\label{fig:bh_init}
\end{figure*}

\begin{table*}[t]
\begin{center}
{\footnotesize
\begin{tabular}{@{\extracolsep{4pt}}lccccc@{}}
\hline
\hline
Basis & State $n$ & Energy gap ($\Delta E_{0n}$) & Dipole moment ($\mu_n$) & Transition dipole moment ($t_{0n}$) & Oscillator strength ($f_{0n}$) \\
\hline
aug-cc-pVDZ & 0 $\;$ (${}^1\Sigma^+$)     & -           &  0.528082(7) & -           & -           \\
            & 1 $\;$ ($\; {}^1\Delta \;$) & 0.216230(3) & -0.18983(3)  & 0.0         & 0.0         \\
            & 2 $\;$ (${}^1\Sigma^+$)     & 0.23727(1)  & -1.4146(5)   & 0.93478(3)  & 0.13822(1)  \\
            & 3 $\;$ (${}^1\Sigma^+$)     & 0.257587(4) & -0.3219(3)   & 0.2102(1)   & 0.007590(9) \\
            & 4 $\;$ (${}^1\Sigma^+$)     & 0.282665(1) &  3.5459(1)   & 0.44725(4)  & 0.037696(7) \\
\hline
aug-cc-pVTZ & 0 $\;$ (${}^1\Sigma^+$)     & -           &  0.54561(2) & -          & -           \\
            & 1 $\;$ ($\; {}^1\Delta \;$) & 0.211482(6) & -0.19271(7) & 0.0        & 0.0         \\
            & 2 $\;$ (${}^1\Sigma^+$)     & 0.238668(8) & -1.2943(5)  & 0.88508(5) & 0.12464(1)  \\
            & 3 $\;$ (${}^1\Sigma^+$)     & 0.253574(4) & -0.4973(6)  & 0.1454(2)  & 0.00358(1)  \\
            & 4 $\;$ (${}^1\Sigma^+$)     & 0.283481(2) &  3.4088(2)  & 0.35740(7) & 0.024141(9) \\
\hline
aug-cc-pVQZ & 0 $\;$ (${}^1\Sigma^+$)     & -          &  0.54914(6) & -         & -          \\
            & 1 $\;$ ($\; {}^1\Delta \;$) & 0.21059(2) & -0.1968(3)  & 0.0       & 0.0        \\
            & 2 $\;$ (${}^1\Sigma^+$)     & 0.23876(3) & -1.268(3)   & 0.8704(3) & 0.1206(1)  \\
            & 3 $\;$ (${}^1\Sigma^+$)     & 0.25261(3) & -0.504(3)   & 0.139(1)  & 0.00327(7) \\
            & 4 $\;$ (${}^1\Sigma^+$)     & 0.28289(1) &  3.2889(9)  & 0.3138(1) & 0.01857(2) \\
\hline
\hline
\end{tabular}
}
\caption{Final converged estimates for the BH molecule at an internuclear distance of $1.2324$\AA. Results are for the five lowest energy states in the $A_1$ irrep of the $C_{2v}$ point group, with $M_S=0$ and $S=\textrm{even}$ quantum numbers. $n=0$ refers to the ground state, $n>1$ to excited states. Numbers in parentheses denote stochastic error, not initiator error. Energy gaps ($\Delta E_{1n}$) were calculated using RDM-based energy estimates, Eq.~(\ref{eq:rdm_energy}). Integrals were generated using the PySCF program\cite{pyscf}.}
\label{tab:bh}
\end{center}
\end{table*}

The calculation of dipole moments provides a more interesting test, due to their greater dependence on more highly-excited determinants and diffuse single-particle orbitals. The relative initiator error is much larger, particularly for certain excited states (i.e. $\mu_2$ and $\mu_3$). The transition dipole moments considered involve transitions from the ground ($n=0$) state to excited ($n>1$) states. Because they always involve the ground state, it is to be expected that they have smaller relative initiator and stochastic error, compared to the corresponding non-transition dipole moment (i.e. $t_{0n}$ compared to $\mu_n$). This expectation is borne out in the results, with initiator and stochastic error in $t_{0n}$ often being $\sim 5$ times smaller than for $\mu_n$. For the calculation of dipole moments from FCIQMC-sampled RDMs, relative stochastic errors are clearly much larger than for energies, and so the use of the semi-stochastic adaptation is of great importance here, whereas its use can be somewhat unnecessary in small ground-state energy calculations.

Clearly, the accurate calculation of dipole moments is more challenging than energies, requiring larger walker populations to obtain similar relative errors. However, this is not uniquely a feature of the initiator approximation in FCIQMC, but is equally true in other approximate methods, where properties such as the dipole moment are far more sensitive to the basis set and quality of the wavefunction than ground state energetics. That we are able to observe systematic converge of these quantities, with respect to a single simulation parameter, is reassuring.

Table~\ref{tab:lih} gives final results for the aug-cc-pV$X$Z basis sets, with $X=2,3,4$. Results in the small $X=2$ basis were fully converged at the smallest walker populations considered, $N_w = 1.25 \times 10^4$, as confirmed by comparison to FCI results from the PySCF program. As expected, dipole moments vary quite substantially with basis set, particularly for the second, third and fourth excited states, demonstrating the importance of large basis sets with diffuse functions. Errors in brackets denote stochastic error bars, not initiator error, which is larger. However, given the careful convergence of initiator error, as shown in Figure~\ref{fig:lih_init}, we expect dipole moments to be converged to around $10^{-3}e a_0$ in most cases, and energies to be converged \emph{substantially} beyond chemical accuracy.

\subsection{BH}

Figure~\ref{fig:bh_init} shows results for BH in the aug-cc-pVTZ basis set and at an internuclear distance of $1.2324$\AA, demonstrating similar initiator convergence plots to those in Figure~\ref{fig:lih_init}. Here, results used between $1.25 \times 10^4$ and $2 \times 10^6$ walkers per simulation. RDM estimators and $E_{\textrm{Trial}}$ were averaged over $5 \times 10^4$ iterations, once convergence was achieved for all states and estimators. Here, instead of using CISD solutions as trial wave functions for $E_{\textrm{Trial}}$, a slightly different approach was used: a ``trial space'' was defined as consisting of the $2 \times 10^3$ most populated configurations across all simulations, once convergence had been approximately reached. Trial wave functions were then obtained as the eigenstates of $\hat{H}$ within this subspace. This is similar to the approach to generate the deterministic space, as described above\cite{Blunt2015}, and allows important basis states to be picked, while allowing an inexpensive calculation to determine each $| \Psi_{\textrm{Trial}}^n \ket$.

Results contain the same features as observed for LiH. Initiator error in the energy estimates are extremely small in all cases, particularly for estimates obtained from contraction of the RDM, and initiator convergence always occurs variationally. Stochastic error bars are larger for $E_{\textrm{RDM}}$, as well as for excited states, but always extremely small. For dipole moments, similar trends also occur. Initiator and stochastic relative errors for the dipole moment are very small for the ground and first excited states ($\mu_0$ and $\mu_1$) and for the corresponding transition dipole moment ($t_{01}$) even at small walker populations. However, results for higher excited states contain larger errors, although we once again observe that errors in $t_{0n}$ are smaller than errors in $\mu_n$ for each $n$, presumably because of the involvement of the ground state, which is well converged at lower walker populations, in each of the transition dipole moments considered.

Table~\ref{tab:bh} shows final results in aug-cc-pV$X$Z basis sets, for $X=2,3,4$. Results for $X=2$ used $2 \times 10^5$ walkers per simulation, while results for $X=3$ and $X=4$ results used $2 \times 10^6$ walkers per simulation. The expected strong dependence of dipole moments on the basis set is once again observed. This is particularly true for the second, third and fourth excited states ($n=2,3,4$). We note that these three states also contained the largest initiator error at small walker populations, as seen in Figure~\ref{fig:bh_init}. This is probably not a coincidence, since the initiator approximation will inevitably result in a poorer description of highly excited regions of the wave function, presumably including excitations into high-energy diffuse functions, which appear important for accurate calculation of dipole moments for these particular states. Despite larger initiator error than for energy estimates, there is still a substantial undersampling of the space here, using $2 \times 10^6$ walkers for a space size of $\sim 7 \times 10^9$ for the aug-cc-pVQZ basis, even for this small molecule, with benefits of Monte Carlo sampling typically increasing with system size.

\subsection{MgO}

\begin{figure*}[t!]
\includegraphics{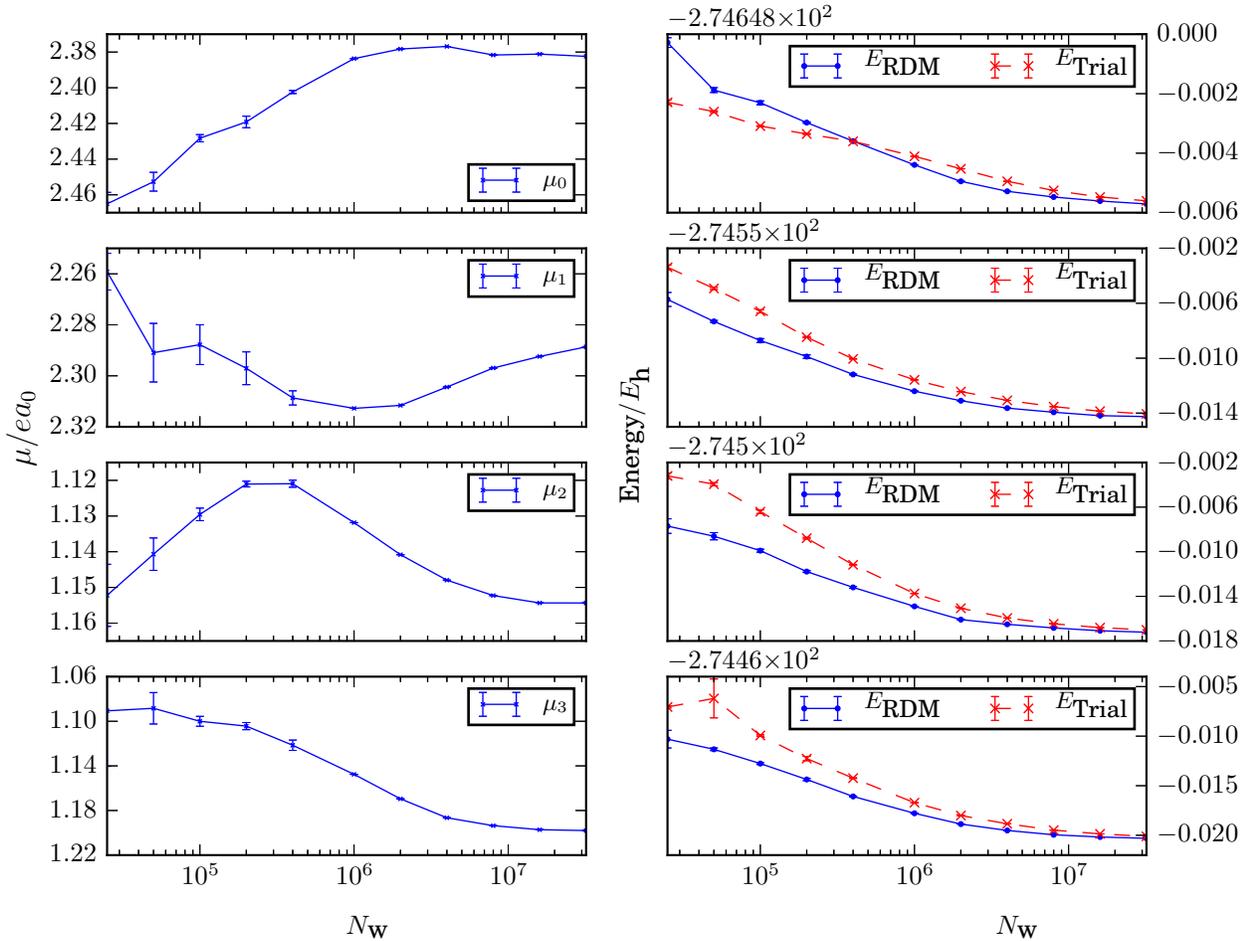}
\caption{Initiator convergence for dipole moments (left) and energies (right), for MgO in an aug-cc-pVDZ basis set, at an internuclear distance of $1.749$\AA, and with 4 core electrons frozen. The four lowest-energy states are considered in the $A_1$ irrep of $C_{2v}$ and with $S=\textrm{even}$ enforced (all ${}^1\Sigma^+$ states). Energies are calculated from both RDM ($E_{\textrm{RDM}}$) and trial wave function ($E_{\textrm{Trial}}$) based estimates, and become equal to good accuracy at large walker number, $N_w$. Dipole moments appear mostly converged at $N_w=3.2 \times 10^7$, except for $\mu_1$. Error bars are only available for $N_w < 10^6$, but are small by this point and should only decrease in magnitude for larger walker populations.}
\label{fig:mgo_init}
\end{figure*}

\begin{table*}[t]
\begin{center}
{\footnotesize
\begin{tabular}{@{\extracolsep{4pt}}c|ccc|ccc@{}}
\hline
\hline
State $n$  & \multicolumn{3}{c|}{ Energy/$E_{\textrm{h}}$ } & \multicolumn{3}{c}{ Dipole moment ($\mu_n$) /$ea_0$ } \\
\hline
 & CCSD & CCSDT & FCIQMC & CCSD & CCSDT & FCIQMC \\
\hline
0 $\;$ (${}^1\Sigma^+$) $\;$ &  -274.632  &  -274.651  &  -274.654 &  2.590  &  2.398  &  2.382 \\
1 $\;$ (${}^1\Sigma^+$) $\;$ &  -274.531  &  -274.559  &  -274.564 &  1.811  &  2.008  &  2.289 \\
2 $\;$ (${}^1\Sigma^+$) $\;$ &  -274.480  &  -274.514  &  -274.517 &  0.297  &  0.847  &  1.154 \\
3 $\;$ (${}^1\Sigma^+$) $\;$ &  -274.440  &  -274.478  &  -274.480 & -0.366  &  0.529  &  1.198 \\
\hline
\hline
\end{tabular}
}
\caption{Energies and dipole moments for MgO in an aug-cc-pVDZ basis set, at an internuclear distance of $1.749$\AA, and with 4 core electrons frozen at the Hartree--Fock level. The four lowest-energy states are considered in the $A_1$ irrep of $C_{2v}$ and with $S=\textrm{even}$ enforced (all ${}^1\Sigma^+$ states). Error bars on FCIQMC results are not given, but are smaller than the order to which results are presented. FCIQMC energies are taken from the RDM-based estimates, $E_{\textrm{RDM}}$. CCSD and CCSDT values were obtained from NWChem\cite{NWChem}.}
\label{tab:mgo}
\end{center}
\end{table*}

To study a more challenging problem, we consider the calculation of energies and dipole moments for the MgO molecule, at its ground state equilibrium separation of $1.749$\AA, and with 4 core electrons frozen at the Hartree--Fock level. Thus, a total of 16 electrons are correlated in $48$ spatial orbitals. Enforcing $M_s=0$, using the $A_1$ irrep of the $C_{2v}$ point group, and working with time-reversal symmetrized functions\cite{Smeyers1973} (to enforce $S=\textrm{even}$), results in a space size of roughly $1.8 \times 10^{16}$ basis functions. This is a large space, particularly given the challenges of converging initiator error in excited-state dipole moments, as seen already.

Figure~\ref{fig:mgo_init} presents initiator convergence for walker populations (per state and per replica), $N_w$, ranging from $2.5 \times 10^4$ to $3.2 \times 10^7$. The ground state and first three excited states are calculated. For $N_w \le 4 \times 10^5$, error bars are calculated by averaging over 5 repeated calculations with varying RNG seeds. Due to the expensive nature of calculations, repeats were not performed for $N_w > 4 \times 10^5$, and so error bars were not obtained. However, these error bars should mostly only decrease with increasing $N_w$, and are already small at $N_w = 4 \times 10^5$. Therefore, at the largest walker populations considered, stochastic error should be much smaller than initiator error.

Initiator profiles of both $E_{\textrm{RDM}}$ and $E_{\textrm{Trial}}$ estimators are presented in Figure~\ref{fig:mgo_init}. At convergence, these should clearly become equal. By $N_w = 3.2 \times 10^7$, this is the case to much better than $1$m$E_\textrm{h}$ accuracy. As previously found, convergence is monotonic in all cases and $E_{\textrm{RDM}}$ usually results in smaller initiator error.

Convergence of dipole moments is also shown. Here, relative initiator error is once again larger than for energies, and convergence is non-monotonic. Because of this non-monotonic behavior, combined with the challenging nature of the system, our confidence in the accurate convergence of these values is somewhat less than for LiH and BH results. We cannot rule out the possibility of sudden further convergence at higher $N_w$ values. However we believe any significant deviations unlikely, although it is clear that $\mu_1$ in particular is not fully converged on the scale shown.

Table~\ref{tab:mgo} presents FCIQMC energies and dipole moments, using $N_w = 3.2 \times 10^7$, and with energies taken from the $E_{\textrm{RDM}}$ estimator. For comparison, coupled cluster results are shown, using both singles and doubles (CCSD) and singles, doubles and triples (CCSDT). These values were calculated using NWChem package\cite{NWChem}, with the equation-of-motion (EOM-CCSD and EOM-CCSDT) variants used for excited states. As expected, energies obtained from CCSDT are accurate compared to FCIQMC values, even for excited states. Meanwhile, dipole moments show greater differences, particularly for the $n=3$ state. For this state, EOM-CCSD and EOM-CCSDT values also greatly differ, with a flipped dipole moment resulting from EOM-CCSD. These results are consistent with those observed in FCIQMC in regions of large initiator error, that the relative error in dipole moments is much greater than in energies. We again expect that this is primarily due to the increased dependence on highly-excited determinants, and such configurations have particularly large amplitudes in excited states. CCSD and CCSDT appear to be unable to describe the wave function with sufficient accuracy in this region of configuration space, for this system, and for these challenging states.

\section{Conclusion}
\label{sec:conclusion}

We have presented the first calculation of excited-state density matrices and transition density matrices using the excited-state FCIQMC method, and subsequently performed accurate RDM-based calculations of energies and dipole moments in small heteronuclear molecules. Initiator convergence has been investigated in some detail for various estimators and excited states, giving further insight into the initiator approach in more general situations than previously considered.

To allow the storage of multiple density matrices in large single-particle basis sets, we have described and implemented an efficient algorithm to store sampled density matrix elements in a sparse format, one which automatically makes full use of symmetries and is efficiently distributed in a massively-parallel simulation. While currently only used for sampling one and two body density matrices, this approach will trivially allow efficient storage of higher body density matrices too, which could be used to calculate important entanglement entropy measures, among many other potential applications.

Although the applications here were to small systems, the calculation of RDMs in FCIQMC will perhaps be key in allowing this QMC approach to be extended to significantly larger problems: this QMC approach has recently been used to develop stochastic versions of the complete active space self-consistent field (CASSCF) method\cite{Thomas2015_3, Manni2016}, in which 1- and 2-RDMs are required to perform the orbital optimization step. RDMs are also required in the explicitly correlated $[2]_{R12}$ approach of Torheyden and co-workers\cite{Torheyden2009,Kong:JCP135-214105}, which can be used with FCIQMC to greatly reduce basis set errors\cite{Kersten2016}, also allowing larger systems to be studied for a given computational cost and accuracy. Given the clear importance of density matrices in FCIQMC currently, we hope that these developments and insights will be of great value going forward.

\section{Acknowledgments}

N.S.B. acknowledges support during this work from Trinity College, Cambridge.
G.H.B. gratefully acknowledges the Royal Society for funding via a university research fellowship, and the support of the Air Force Office of Scientific Research (AFOSR) through grant award no. FA9550-16-1-0256. A.A. acknowledges support by the EPSRC under grant no. EP/J003867/1.
We also thank Qiming Sun for helpful conversations regarding the PySCF program.

\end{document}